\documentclass[useAMS]{mn2e}
\usepackage{psfig,epsfig}
\usepackage{graphicx}
\usepackage{amssymb}
\usepackage{epstopdf}

\begin{document}
\def\simlt{\mathrel{\rlap{\lower 3pt\hbox{$\sim$}}
        \raise 2.0pt\hbox{$<$}}}
\def\simgt{\mathrel{\rlap{\lower 3pt\hbox{$\sim$}}
        \raise 2.0pt\hbox{$>$}}}

\title[The clustering properties of high-redshift passive galaxies]{The clustering properties of high-redshift passive galaxies}

\author[Manuela Magliocchetti et al.]
{\parbox[t]\textwidth{M. Magliocchetti$^{1}$,
P.Santini$^{2}$, E.Merlin$^{2}$, 
L.Pentericci$^{2}$}\\
 \\
{\tt $^1$ INAF-IAPS, Via Fosso del Cavaliere 100, 00133 Roma,
  Italy}\\
{\tt $^2$ INAF - Osservatorio Astronomico di Roma, Via di Frascati 33, 00078, Monteporzio Catone, Italy} }
\maketitle
\begin{abstract}
We investigate the clustering properties of $3< z < 5$ candidate passive galaxies from the Merlin et al. (2019) sample residing in the GOODS-North (35 sources) and GOODS-South (33 sources) fields. Within the large uncertainties due to the paucity of sources we do not detect clustering signal in GOODS-North, while this is present in GOODS-South, highlighting the importance of the effects of cosmic variance. The estimated correlation length in GOODS-South is $r_0=12^{+4}_{-5}$~Mpc, while the estimated minimum mass for a halo capable to host one of such high-redshift quenched galaxies is log$_{10}\left(M_{\rm min}/M_\odot\right) =13.0\pm 0.3$, once also the constraints from their space density are taken into account. Both values are compatible with the results from GOODS-North. 
Putting the above findings in a cosmological context, these suggest no evolution of the dark matter content of the hosts of passive galaxies during the past 12.5 Gyr, i.e. during more than 90\% of the age of the Universe. We discuss possible scenarios for the observed trend.

 \end{abstract}

\section{Introduction}
The advent of the James Webb Space Telescope (JWST) is quickly revolutionizing our view on the formation and evolution of galaxies. In particular, for what concerns passive galaxies, it is becoming evident that more than expected exist at epochs as early as $z\sim 5$ (e.g. Carnall et al. 2023; Valentino et al. 2023; Long et al. 2023), with one of them recently detected at $z\sim 7$ (Looser et al. 2023). This is connected to the so-far unforeseen rapid growth of early seeds into massive galaxies with log$_{10}\left(M_*/M_\odot\right)=10-11$ during the latter half of the first billion years, from $6 < z < 10$ (Labbe et al. 2023). This finding has led to a burst of activity aimed at assessing whether the existence and abundance of such objects poses a threat for current models of structure formation (e.g. Menci et al. 2022; Lovell et al. 2023). Surely enough, JWST observations have led to a revision of models for galaxy formation at least in the primeval universe, since the existence of passive galaxies as early as less than 1 Gyr after the Big Bang requires a very rapid ($\sim 200$~Myr - Carnall et al. 2023a) phase of star-formation followed by a fast quenching in their parent population at even earlier epochs. 

Due to its relatively limited field of view, one element that JWST observations are currently still missing is the investigation of the spatial properties of high-redshift galaxies in order to help putting their existence in a cosmological context. We aim at filling this gap by presenting the clustering properties of a homogeneous population of passive galaxies selected within $3<z<5$ in the GOODS-North and GOODS-South fields over a total area of $\sim 350$ arcmin$^2$ by the work of Merlin et al. (2019) (M19 hereafter). 

The layout of the paper is as follows: \S2 briefly describes the Merlin et al. (2019) sample and shows our estimates for the angular two-point correlation function $w(\theta)$ in both fields, while \S3 presents the results concerning physical quantities such as the correlation length and minimum halo mass derived from our clustering measurements. \S4 discusses the results in a cosmological context of structure growth and summarises our conclusions. Throughout this paper we assume a $\Lambda$CDM cosmology with $H_0=70 \: \rm km\:s^{-1}\: Mpc^{-1}$ ($h=0.7$), $\Omega_0=0.3$,  $\Omega_\Lambda=0.7$ and $\sigma^m_8=0.8$.

\section{The Data}

\begin{figure*}
\vspace{-1cm}  
\includegraphics[scale=0.4,clip,viewport=-20 160 600 760]{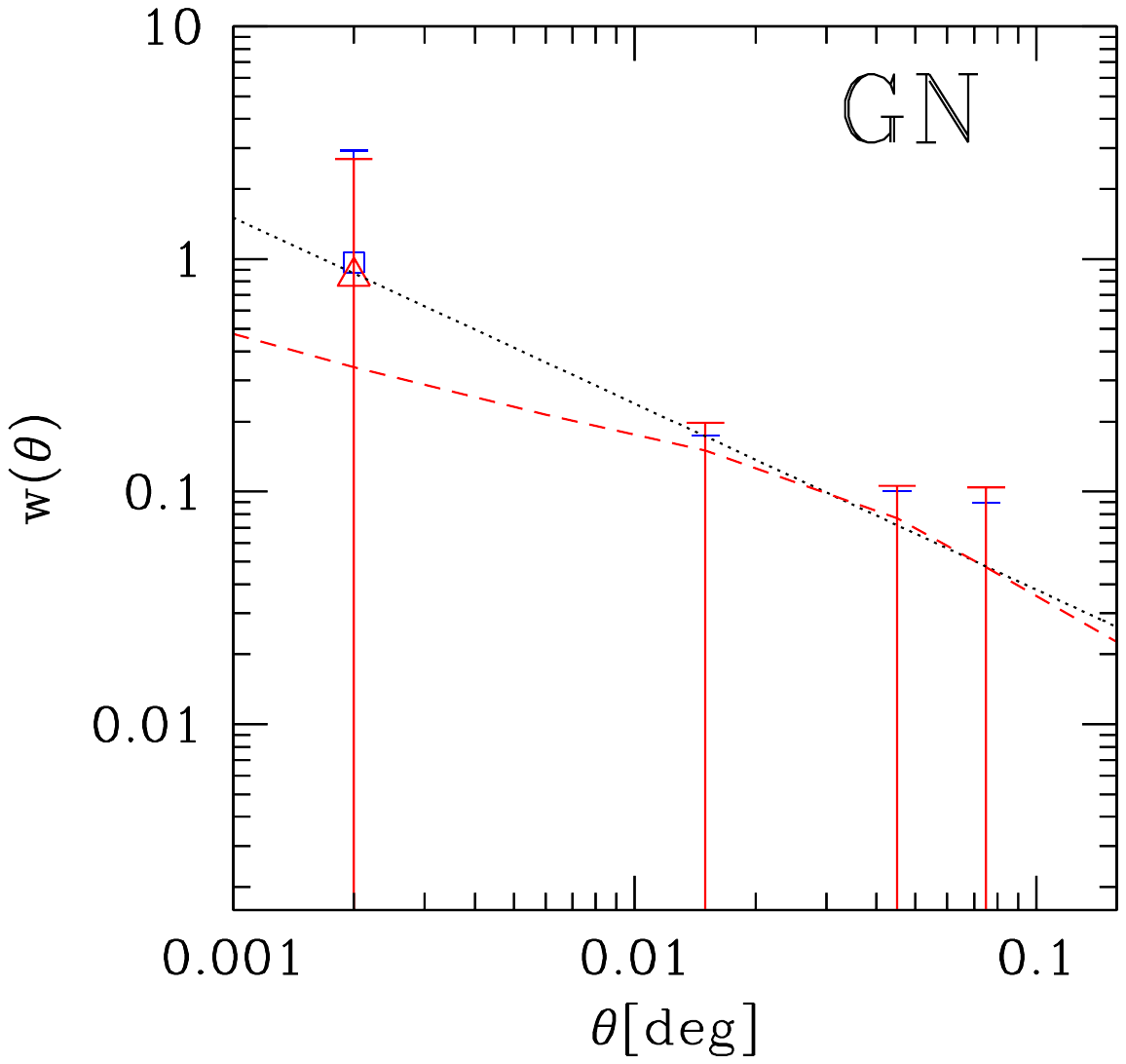}
\includegraphics[scale=0.4,clip,viewport=-20 160 600 760]{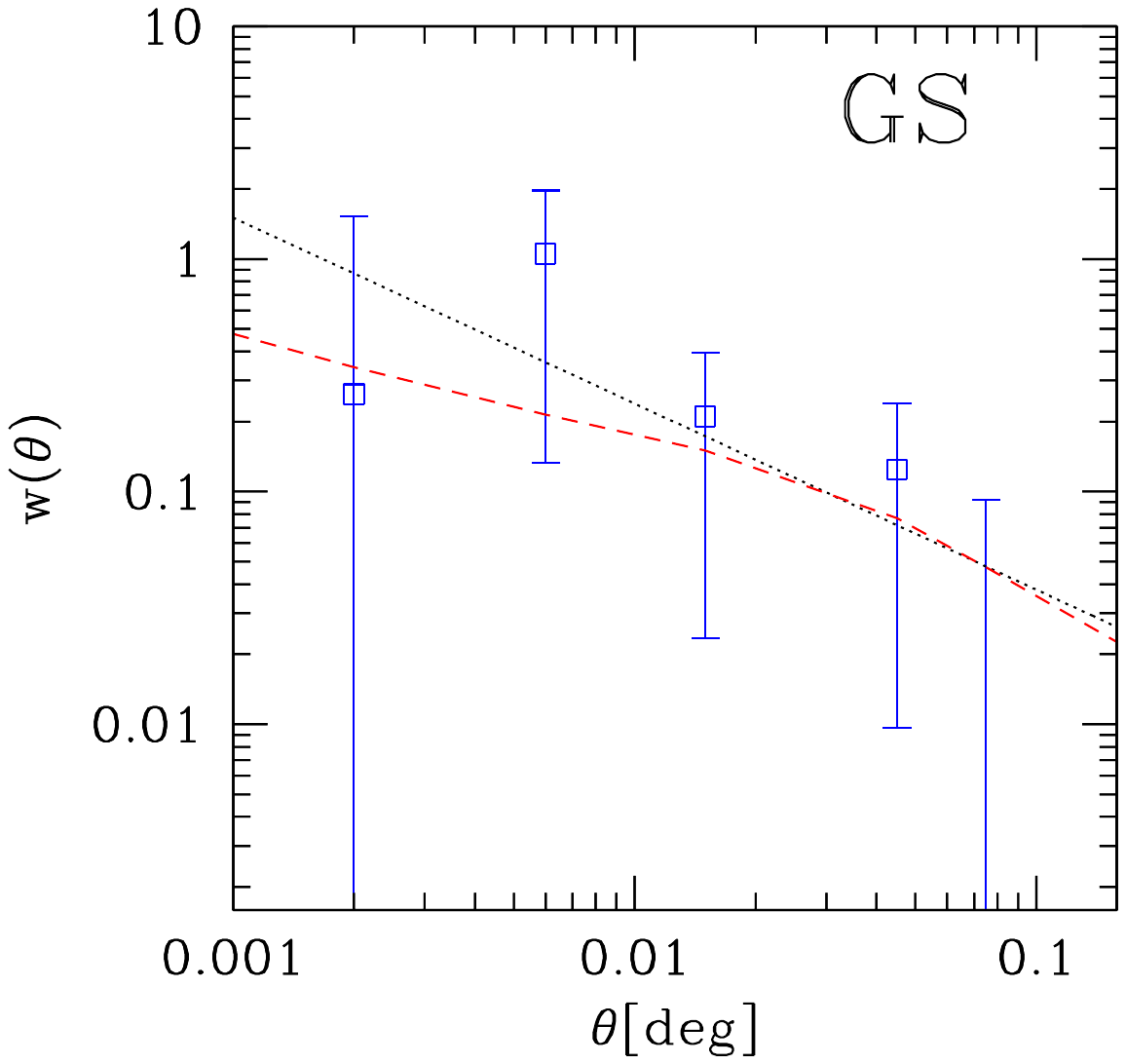}
\caption{Angular two-point correlation function $w(\theta)$ for $3< z< 5$ passive galaxies in GOODS-North (GN - left-hand panel) and GOODS-South (GS - right-hand panel). Errorbars indicate Poissonian uncertainties. In the left-hand panel the blue squares and associated errors correspond to the whole GN sample, while the red triangles to a subsample obtained by removing the peripheral sources in the field. In both panels the curves correspond to the best fits to the observed GS $w(\theta)$: the dotted straight line is for a power-law of the form $\left(r/r_0\right)^{-1.8}$ with $r_0=12$~Mpc, while the dashed curve is for a physical model whereby galaxies are hosted within dark matter halos more massive than log$_{10}\left(M_{\rm {min}}/M_\odot\right)=13$ (see text for details).
\label{fig:w_theta}}
\end{figure*}

We consider the sample of $z>3$ passive galaxy candidates photometrically selected by M19 over the CANDELS area. The selection is based on SED fitting adopting a star formation history (SFH) characterized by a period of constant star formation followed by an instantaneous suppression. Such a SFH has been demonstrated to be more appropriate for mimicking the rapid timescales involved in galaxy quenching in the early Universe (Merlin et al. 2018). Galaxies are pre-selected to have $H$$<$27, 1$\sigma$ detection in $K$, IRAC1 and IRAC2 bands and photometric redshift $>$3. Besides requiring a passive best-fit solution (SFR$=$0), only the candidates with $>$30\% associated probability of being passive and secondary star-forming solutions with probabilities not higher than 5\% are included in the selection. These requirements make the selection very robust, as demonstrated by the lack of  on-going star formation probed by available ALMA observations (Santini et al. 2019; 2021).

Due to their higher completeness in the selection of high-redshift passive galaxies with respect to the other CANDELS fields (COSMOS, EGS and UDS, cfr Section 5.1 and Figures 2 and 6 of M19), for the present analysis we only consider GOODS-North (hereafter GN) and GOODS-South (hereafter GS). In the redshift range $3<z<5$ there are respectively 35 passive galaxies in GN and 33 in GS.
Their projected distribution onto the sky is presented in Figure 1 of M19. From these samples we estimated the angular two-point correlation function $w(\theta)$ which gives the excess probability, with respect to a random Poisson distribution, of finding two sources in the solid angles $\delta\Omega_1$ $\delta\Omega_2$ separated by an angle $\theta$. In practical terms, $w(\theta)$ is obtained by comparing the observed source distribution with a catalogue of randomly distributed objects subject to the same mask constraints as the real data. We used the estimator (Hamilton 1993)
\begin{eqnarray}
w(\theta) = 4\times \frac{DD\cdot RR}{(DR)^2} -1, 
\label{eq:xiest}
\end{eqnarray}
where $DD$, $RR$ and $DR$ are the number of data-data, random-random and data-random pairs separated by a distance $\theta$. 

Given the roughly constant photometric coverage along the two fields, we have then simply generated two random catalogues (one for GN and one for GS) covering the whole surveyed areas with twenty times as many objects as the real datasets and estimated $w(\theta)$  in the angular range $10^{-3}\simlt \theta \simlt 0.1$ degrees, where the upper limit corresponds to about half of the maximum scale probed by the surveys.  
Furthermore, since GN includes a handful of sources with a peculiar location right at the edges of the field (M19), we have also recalculated $w(\theta)$ in GN for a subsample of 28 galaxies which exclude the border area.  The results are presented in Figure~\ref{fig:w_theta}, whereby in the left-hand panel the blue squares with associated (Poissonian) uncertainties are for the whole $3<z<5$ GN, while the red triangles are for the aforementioned subsample. As it is possible to appreciate, within the large uncertainties due to the paucity of sources, no clustering signal is observed in GN in both cases.  Positive clustering signal is instead observed in GS (right-hand panel of Figure~\ref{fig:w_theta}). These discrepant results can be explained with a combination of two effects: 1- cosmic variance, especially important in small fields such as those analysed here and 2- the fact that GN is more incomplete with respect to GS (especially in the four IRAC bands - cfr Figure 2 of M19) when it comes to the selection of high-z passive galaxies. This last point implicates that we expect the $w(\theta)$ measurements in GS to be more reliable than those in GN and therefore in the following we will only concentrate on them. However, since from the current observations it is not possible to disentangle the effect of cosmic variance from that arising from incomplete selection, throughout the paper we will still consider the results from GN in order to make sure that estimates arising from GS do not contradict them. For the time being we note that, within the large uncertainties, measurements of $w(\theta)$ in the two fields are compatible with each other.\\
To parametrize the clustering signal observed in GS, we assume for the angular two-point correlation function the standard power-law form $w(\theta)=A\theta^{1-\gamma}$, 
where the parameters $A$ and $\gamma$ can be obtained from a least-squares fit to the data.  Given the large errors on $w$,  we chose to fix $\gamma$ to the standard value $\gamma=1.8$.
The small area of GS introduces a negative bias through the integral constraint $\int w(\theta) d\Omega = 0 $. We allow for this by fitting to $ A \theta^{1-\gamma} -C $, where $C = 1.78 A$.

The best fit to the GS data is obtained for an amplitude $A=(6\pm 4)\cdot 10^{-3}$ and is represented by the dotted line in both the left-hand and right-hand panels of Figure {\ref{fig:w_theta}}, showing that 
 (within the large errors) it also compatible with the GN results.  We stress that in all cases the quoted errors correspond to Poissonian statistics. We are aware that this leads to somewhat smaller uncertainties than what expected (even though the issue is less problematic in small fields, e.g. Willumsen, Freudling \& Da Costa 1997) and this is mainly the reason for keeping $w(\theta)$ from GN as a testbed for our results. A further discussion on this topic and on the reliability of our results is presented in \S3.

\section{Physical parameters}

\begin{figure} 
\vspace{-1 cm}  
\includegraphics[scale=0.4,clip,viewport=-20 190 600 760]{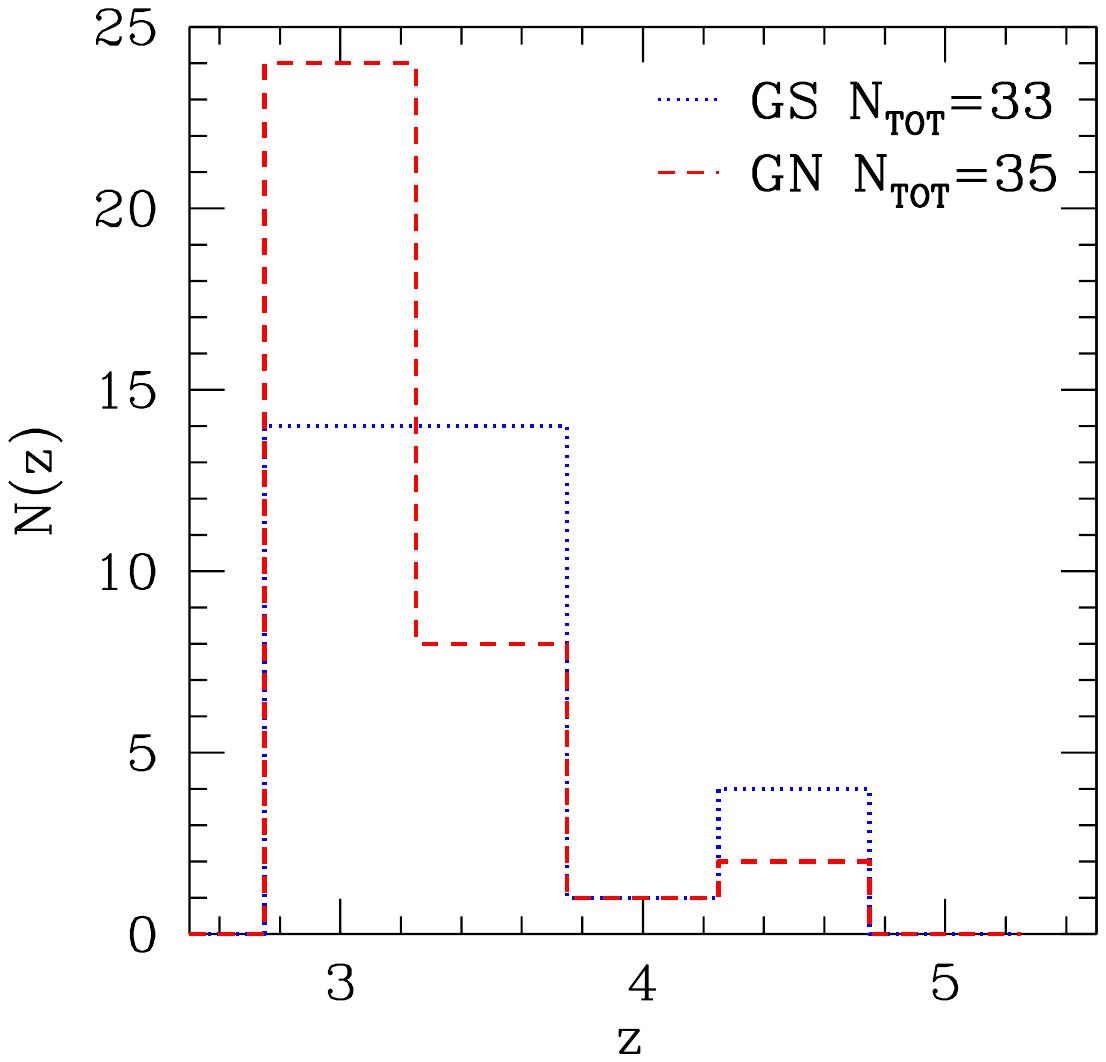}
\caption{Redshift distributions for $3 < z < 5$ passive galaxies in GOODS-South (GS) and GOODS-North (GN).
\label{fig:N_z}}
\end{figure}

The standard way of relating the angular two-point correlation function $w(\theta)$ to the spatial two-point correlation function $\xi(r,z)$ is by means of the relativistic Limber equation (Peebles
1980; Magliocchetti \& Maddox 1999) which makes use of the knowledge of the redshift distribution $N(z)$ of the sources under exam.

We adopt a spatial correlation function of the form $\xi(r,z)=(r/r_0)^{-\gamma}$, where we assume the correlation length $r_0$ not to vary in the considered redshift interval (assumption justified by the fact that the $z=[3-5]$ range only spans less than 1 Gyr) and -- in agreement with \S2 --  further fix the slope to $\gamma=1.8$. For the adopted cosmology and the redshift distribution presented in Figure~{\ref{fig:N_z}} in the case of GS we obtain $r_0=12^{+4}_{-5}$~Mpc (comoving). This is in full agreement with the results from e.g. McCracken et al. (2010) and Hartley et al. (2013) for their populations of intermediate redshift ($z\sim[1-3]$ in the first case, $z\sim [2-3]$ in the second) passive galaxies. Note that in the process of data-fitting, fixing $\gamma$ implies not considering the covariance between the amplitude of the correlation function and its slope. In general, higher values of $\gamma$ correspond to lower values of $r_0$.

A more meaningful way to derive physical quantities from clustering estimates is by means of the Halo Bias approach (Mo \& White 1996; Sheth \& Tormen 1999) 
which describes the observed clustering signal in terms of the product between the two-point correlation function of the underlying dark matter distribution $\xi_ {\it dm}(r,z)$ and the square of the so-called 
bias function $b(M_{\rm min},z)$,  which at a given redshift solely depends on the minimum mass $M_{\rm min}$ of the dark matter halos in which the detected astrophysical objects reside: 
\begin{equation}
\xi(r,z)=\xi_ {\it dm}(r,z) \cdot b^2(M_{\rm min},z).
\label{eq:bias}
\end{equation}
More sophisticated models have been introduced in the recent years to provide a more realistic description of the observed clustering of extragalactic sources (Halo Occupation Model or HOM, e.g. Scoccimarro et al. 2001).  
These relax the assumption, implicit in the Halo Bias model,  of having a one-to-one correspondence between a galaxy and its  halo.
 Such an analysis is unfortunately not possible with our dataset (and with all those not including a large enough number of sources, as is the case of most high-redshift surveys) due to a relatively poor signal-to-noise ratio. For the scopes of the present work, it just suffices noticing that in presence of multiple halo occupancy, the values for $M_{\rm min}$ found with the HOM approach will be (slightly, of the order of 0.5 dex at least up to $z\sim 2$ - cfr Magliocchetti et al. 2008) lower than those obtained via the Halo Bias model.

Having made this necessary digression, we stress that the Halo Bias approach is anyhow valid as it provides estimates of a physical quantity which is the mass of the parent halo where the galaxies reside as opposed to that of the galactic sub-halos. We can then use it and obtain the theoretical angular two-point correlation function $w(\theta)_{\rm th}$ predicted by this model 
starting from eq. (\ref{eq:bias}) and projecting it once again by following the Limber equation with the GS $N(z)$ provided in Figure \ref{fig:N_z}. The (linear and non-linear) dark matter correlation function $\xi_ {\it dm}(r,z)$ was computed by following Peacock \& Dodds (1996), while the bias function $b(M_{\rm min},z)$ was obtained from the Sheth \& Tormen (1999) prescriptions. Fitting to the observed GS $w(\theta)$, we obtain a minimum mass for a dark matter halo capable of hosting a $3<z<5$ passive galaxy of log$_{10}\left(M_{\rm{min}}/M_\odot\right)=13.0^{+0.3}_{-1.2}$. The corresponding best fit $w(\theta)_{\rm th}$ is shown in both panels of Figure \ref{fig:w_theta} by the dashed curve, and once again indicates that the result for GS is also consistent within the large uncertainties with the observed $w(\theta)$ in GN.

As it is possible to appreciate, the errors associated to $M_{\rm min}$ are highly asymmetric. This is due to the fact that at large halo masses, even small increments of $M_{\rm{min}}$ produce large positive variations of the bias function in eq. (\ref{eq:bias}) and consequently much larger amplitudes for the expected $w(\theta)_{\rm th}$. Indeed, if we envisage a very extreme case whereby uncertainties on the observed GS $w(\theta)$ are as large as twice those estimated with Poissonian statistics in \S2, the upper limit associated to log$_{10}\left(M_{\rm{min}}/M_\odot\right)$ would at most be 13.5 (instead of 13.3 as found before). 
This ensures that the upper constraint we were able to put on $M_{\rm{min}}$ in the case of $3<z< 5$ passive galaxies is solid and that we confidently do not expect such objects to start appearing within halos of masses larger than $\sim 10^{13.3}-10^{13.4} M\odot$.

Further useful constraints on halo masses come from the observed number density of our passive galaxies. M19 estimate a (corrected for incompleteness) value in GS of 
$(4.4\pm 0.8)\cdot 10^{-5}$ Mpc$^{-3}$. If we then assume, as we have done throughout the paper, one galaxy per dark matter halo and integrate the Sheth \& Tormen (1999) mass function in the considered redshift range over all halos with masses larger than some minimum value we have that, in order to match the observed abundances, halos have to be more massive than $M_{\rm min} = 10^{12.8\pm 0.1}M_\odot$.  
This value is in excellent agreement with the results from clustering measurements, and puts a stringent constraint on the allowed minimum halo mass to be $M_{\rm min} =10^{12.7}M_\odot$.  The previous finding is also consistent with the relatively large stellar masses (mean value $<M_*>=10^{10.4\pm 0.5} M_\odot$) estimated for our sample, since it ensures a ratio between dark and stellar matter of about $10^{2.3}$, in agreement with recent results for early-type galaxies (e.g. Cannarozzo, Sonnenfeld \& Nipoti 2020). Lower halo masses would instead lead to ratios below $10^2$, overshooting the observed scaling relations. It therefore follows that if we combine estimates from clustering and space density we obtain an allowed range for the minimum mass of a halo capable to host at least one high-redshift passive galaxy of log$_{10}\left(M_{\rm min}/M_\odot\right) =13.0\pm 0.3$.

\section{Discussion}


\begin{figure}
\vspace{-1cm}  
\includegraphics[scale=0.4,clip,viewport=-20 160 600 760]{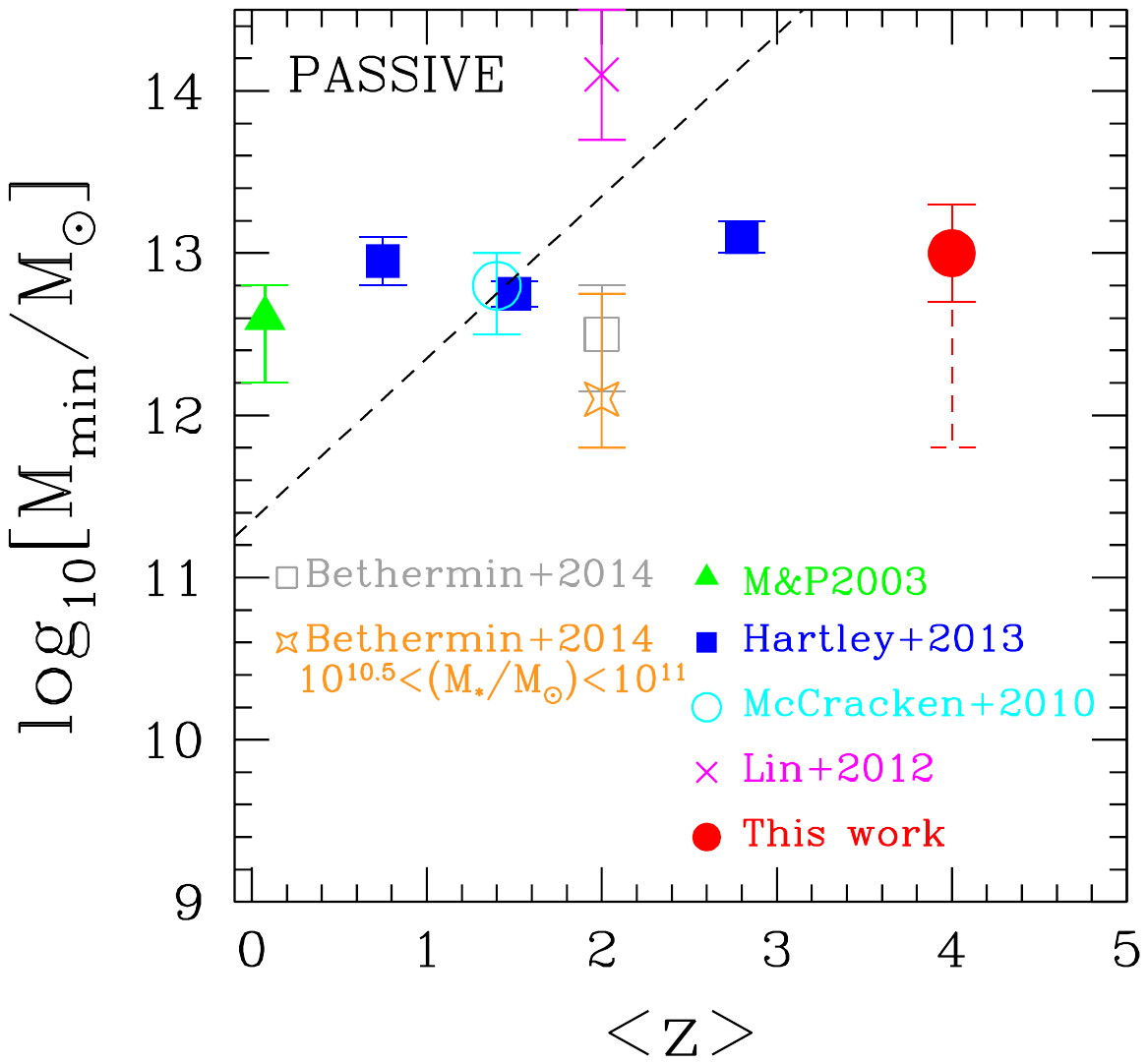}
\caption{Variation with redshift of the minimum halo mass for passive galaxies up to $z\sim 5$. Different symbols correspond to different samples (M\&P2003 stands for Magliocchetti \& Porciani 2003). The dashed line represents the best fit to the data obtained by Magliocchetti et al. (2014) for the cosmological evolution of the minimum mass of halos hosting $z\simlt 3$ star-forming galaxies with enhanced activity ($SFR\simgt 20 \:M_\odot$/yr - see text for details). 
\label{fig:mass}}
\end{figure}

Our results can be put in context with those available from the literature. Magliocchetti \& Porciani (2003) estimate the minimum halo mass for 36,318 spectroscopically-selected early-type galaxies with  $z\simlt 0.1$ from the 2dF Galaxy Redshift Survey (Colless et al. 2001). Despite using a different approach from that adopted in this work, 
they obtain the value log$_{10}\left(M_{\rm min}/M_\odot\right)=12.6^{+0.1}_{-0.5}$.  

At slightly higher redshifts McCracken et al. (2010) investigate the clustering properties of 3,931 BzK-selected passive galaxies in the $z\sim [1-2]$ redshift range within the COSMOS field. In this case we have derived minimum halo masses from their reported bias values obtained at large ($\theta\simgt 0.01$ deg) angular scales, i.e. before the multiple occupancy of the halos provokes a steepening of the observed $w(\theta)$. In order to do so, we have considered the Sheth \& Tormen (1999) bias function at the mean redshift of the McCracken et al. (2010) sample under the assumption (verified at $\theta>0.01$ deg, i.e. in the regime of the two-halo term) of one galaxy per halo. The derived minimum mass is log$_{10}\left(M_{\rm min}/M_\odot\right)=12.8^{+0.2}_{-0.3}$.

At a median redshift of $z\sim 2$  Bethermin et al. (2014) analyse 2,821 passive galaxies within COSMOS selected with the BzK colour technique. Under the hypothesis of one galaxy per dark matter halo, these authors find average masses log$_{10}\left(<M_{\rm halo}>/M_\odot\right)=12.70^{+0.30}_{-0.35}$ for the whole sample and log$_{10}\left(<M_{\rm halo}>/M_\odot\right)=12.30^{+0.65}_{-0.30}$ for a restricted subsample of galaxies with stellar masses $M_*=10^{10.5}-10^{11} M_\odot$, more similar to our objects (cfr \S3). For a Sheth \& Tormen (1999) mass function this implies values for the minimum halo mass respectively of log$_{10}\left(M_{\rm min}/M_\odot\right)=12.50^{+0.30}_{-0.35}$ and log$_{10}\left(M_{\rm min}/M_\odot\right)=12.10^{+0.65}_{-0.30}$. Also the work of Lin et al. (2012) selects passive galaxies with stellar masses comparable to those of our sample at $z\sim 2$ via the BzK technique. These authors find 44 such objects in GN for which they estimate a minimum halo mass log$_{10}\left(M_{\rm min}/M_\odot\right)=14.1\pm 0.4$, again under the hypothesis of one galaxy per halo. 

Hartley et al. (2013) instead investigate the clustering of passive galaxies selected via a technique which combines $UVJ$ colours and specific star-formation rates in the UDS field up to $z\sim 3.5$. They  also consider a one-to-one relation between galaxies and halos. By averaging up the results provided in their Figure 8 for different stellar mass ranges (which cover that of our sample) and redshifts, we obtain log$_{10}\left(M_{\rm min}/M_\odot\right)=13.1\pm 0.1$ at a median redshift 
$<z>\sim 2.8$, log$_{10}\left(M_{\rm min}/M_\odot\right)=12.75\pm 0.08$ at $<z>\sim 1.5$ and log$_{10}\left(M_{\rm min}/M_\odot\right)=12.95\pm 0.15$ at $<z>\sim 0.75$.

The variation of  $M_{\rm min}$ with cosmic epoch in the case of passive galaxies is illustrated in Figure~\ref{fig:mass}, where different symbols and colours represent the different results discussed so far. Our work is indicated by the filled (red) circle: the dashed errorbar represents the uncertainties associated to clustering measurements only, while the solid one also includes the constraints from space density. As it is possible to appreciate, except for the point corresponding to the Lin et al. (2012) work, all estimates are compatible with each other and show no variation with cosmic time in the last $\sim 12.5$ Gyr. We note that the absence of scatter or variations in the trend observed in Figure~\ref{fig:mass} also implies that the different methods adopted to identify passive galaxies (SED-fitting, spectral investigation, BzK technique, specific star-formation rates) are substantially consistent with each other and select the same underlying population.

More works have estimated halo masses for passive galaxies up to redshifts $z\sim 3$ (e.g. Tinker et al. 2013; McCracken et al. 2015; Cowley et al. 2019). A straightforward comparison with our finding results difficult because these all use the full HOM approach as opposed to the Halo Bias model adopted here. In spite of this, also the aforementioned works point towards a general constancy of the halo masses of passive galaxies throughout the cosmic epoch corresponding to $z\sim 0-3$. Furthermore, in stellar mass ranges comparable to that of our sample, they all find values log$_{10}\left(M_{\rm min}/M_\odot\right)\simeq 12-12.5$ which, assuming a difference of about 0.5 dex (cfr \S3) between halo mass values as estimated by the two different methods, corresponds in our case to 12.5-13, in excellent agreement with the results presented in Figure~\ref{fig:mass}.

Taking then the plot in Figure~\ref{fig:mass} at face value, this means that during more than 90\% of the age of the Universe the environmental properties of passive galaxies have remained the same, being all hosted within dark matter halos more massive than log$_{10}\left(M_{\rm min}/M_\odot\right)\simeq 12.8-13$ ($\sim 12.3-12.5$ in the case of a full HOM approach). 

For a visual comparison, we also present with a dashed line the result obtained by Magliocchetti et al. (2014) in the case of $z\simlt 3$ galaxies undergoing an episode of moderate-to-intense star-formation activity (Star-Formation Rates, SFR$\simgt 20\: M_\odot/$yr). This indicates strong downsizing, whereby while high-$z$ star-forming galaxies are hosted by very massive, (up to $10^{14}$ $M_\odot$) proto-clusters or cluster-like structures,  in the nearby universe even large ($\simgt 100$ $M_\odot$/yr) SFRs are associated to much less massive, $10^{11}-10^{12}$ $M_\odot$, halos. 


It then appears that passive galaxies and star-forming galaxies behave very differently. But what is the information one can derive on the formation and subsequent evolution of passive galaxies from the clustering results presented in this work?
We can basically envisage two different scenarios. In the first one, early-type galaxies are fully formed and in place already at as early as $z\sim 5$ or even 7 (e.g. Carnall et al. 2023a; Looser et al. 2023) and their evolution simply proceeds in a totally "passive" way, i.e. not only these sources do not any longer form stars from such redshifts down to the local universe, but also their environmental properties remain the same during all probed cosmological epochs. In other words, the early-type galaxies we see today are the very same objects we observe in the early universe without modifications of any sort. 

The second scenario instead foresees a strong connection between high-z passive galaxies and intermediate-z highly star-forming galaxies, whereby the early-type galaxies that form in the very early universe later become the 'seeds' for fast accretion of gas and other galaxies and therefore transform themselves into the star-forming galaxies set at the centres of clusters and proto-clusters we observe at $z\sim 2-3$ in the compilation of results presented by Magliocchetti et al. (2014). In this case these central galaxies keep accreting material and form stars until the gas reservoir is available and end up in the local universe as very massive CD galaxies. This implies that the passive galaxies we observe in the early universe are not the same passive galaxies we encounter at lower redshifts, since the high-redshift ones will be later found in much more massive dark halos having become cluster central galaxies. This also implies that 'typical' (i.e. non CD-like) early-type galaxies continuously form from $z\sim 5$(7) down to the local universe and therefore that on average we expect them to have lower redshifts of formation (i.e. are younger) than the more massive - CD-like - ones. Assuming that our sample has unveiled most of the very first passive galaxies right after their formation, one prediction from this second scenario is that the number density of $z\sim 5$ early-type galaxies is the same as that of local galaxy clusters, which is indeed the case (e.g. Dekel \& Ostriker 1999). Another concordance point would be the masses of the evolved stellar component for the star-forming galaxies in protocluster environments observed at $z\sim 3$, since the very high values reported ($<M_*>\sim 10^{11.3} $ M$_\odot$, e.g. Magliocchetti et al. 2011) would be hard to achieve in the absence of already massive seeds. These could be identified in the primeval passive galaxies discussed in this work.  On the other hand, we note that one difficulty would be the observed consistency of the clustering properties of 'typical' (i.e. non CD-like) passive galaxies throughout all probed times that could only be explained with some kind of cosmic conspiracy (or fine-tuning) whereby, independent of redshift, early-type galaxies only form at a specific value of the mass of the dark halo which hosts them.  We defer the reader to a future paper which will tackle this issue in greater detail.\\

 \noindent
{\bf Acknowledgements}
We wish to thank the anonymous referee for the constructive comments which helped improving the paper. \\

\noindent
{\bf Data Availability}\\
The data underlying this article will be shared on reasonable request to the corresponding author.

\end{document}